# Self-organization of gas bubbles

M.E. Shcherbina, E.V. Barmina, P.G. Kuzmin, and G.A. Shafeev

*Wave Research Center of A.M. Prokhorov General Physics Institute of the Russian Academy of Sciences, 38, Vavilov street, 119991, Moscow, Russian Federation*

**Abstract**

Self-organization of hydrogen bubbles is reported under etching of metallic Aluminum in a weakly basic solution. The ascending gas bubbles drift to the areas with higher density of bubbles due to pressure difference. As a result, ascending bubbles form various stationary structures whose symmetry is determined by the symmetry of the etched area. The process is modeled on the basis of numerical solution of Bernoulli equation.

Liquid flow in the field of gravity is characterized by a number of instabilities, e.g., Rayleigh-Bénard one or Bénard-Marangoni convection. In some cases these instabilities lead to the formation of self-organized structures with Bénard cells being a well-known example. In both cases the liquid remains homogeneous, and the gas that is dissolved in it does not affect its flow. If the liquid contains another phase, e.g., gas bubbles, then their motion may involve it into flow due to viscous interaction at the bubble-liquid interface. In turn, the liquid flow affects the motion of bubbles so there is a feedback between the flows of the liquid and gas bubbles.

We report here on the self-organization of gas bubbles that takes place during the chemical reaction of etching in aqueous solution in the field of gravity. This phenomenon looks quite general and appears in different chemical processes in liquid phase that are accompanied by the emission of gas bubbles. In our case we have studied the self-organization of hydrogen micro-bubbles that are generated during etching of metallic Aluminum in an aqueous solution of ammonia. The latter is a weak base, and its interaction with Al results in hydrogen emission.

Gas bubbles appear during the reaction of laser-treated bulk Aluminum target with aqueous solution of ammonia at 5% concentration. Laser exposure of the target in air leads to the formation of a surface with high specific area as shown in Fig. 1. The exposed areas look dark compared to the initial substrate.

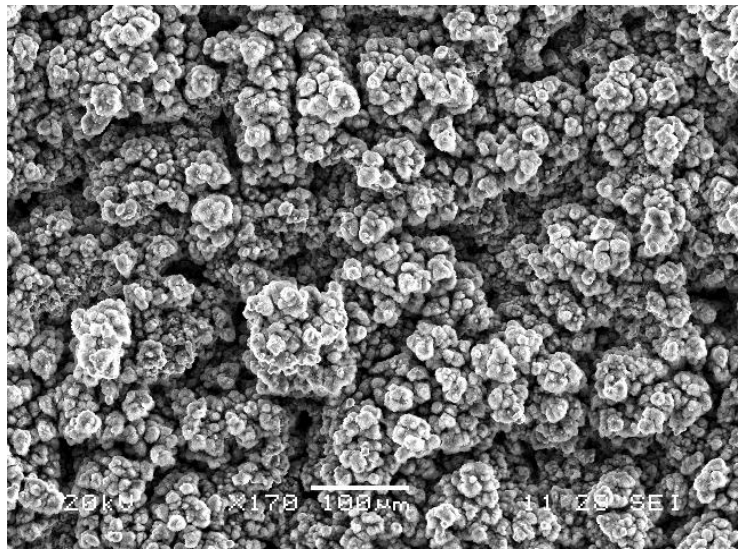

Fig. 1. Scanning Electron Microscope view of the Aluminum surface exposed to radiation of a Nd:YAG laser in air. Wavelength of 1.06 μm, pulse duration of 100 ns. Scale bar denotes 100 μm.

Upon immersion into the etching solution, the micro-bubbles of hydrogen are generated mostly in laser-processed areas of the substrate. After some incubation time of several minutes etching of the aluminum starts and is accompanied by emission of micro-bubbles of hydrogen with estimated diameter between 100 and 200 μm.

The following image shows that the distribution of gas bubbles that is established after some incubation time is not homogeneous (Fig. 2).

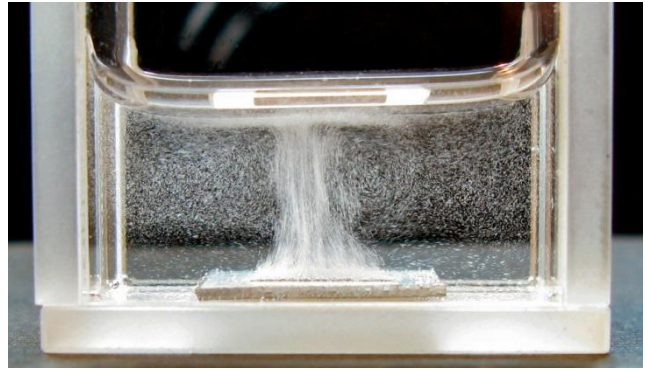

Fig. 2. Side view of hydrogen bubbles distribution under etching of laser-processed Al target. The lateral size of the laser-treated area is 1 cm. Dark line on the target is the edge of the laser-exposed area.

Gas bubbles produced at the periphery of the laser-treated area move to its center with simultaneous ascending motion. They drag the liquid and disappear at the surface, while the liquid flows horizontally reaching the vessel walls and then descends. The length of the tracks of gas bubbles during the shutter exposure (1/8 s) allows estimating the velocities of the liquid flows in various regions of the solution. One can clearly see two vortices in which the bubbles move in a circular way.

The map of bubbles velocity is presented in Fig. 3. One can see that the highest velocity is observed above the center of the etched area (around 5 mm/s).

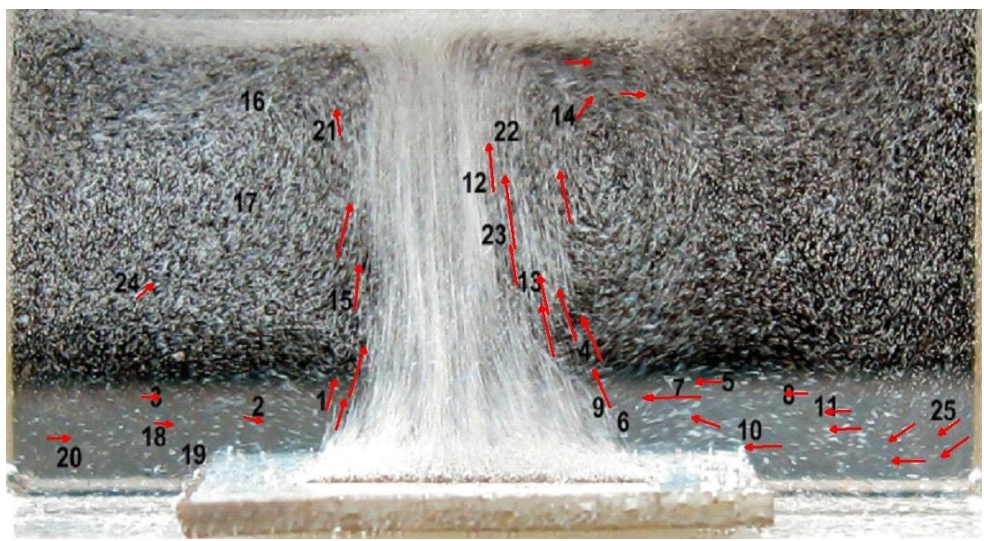

Fig. 3. The map of the bubbles velocities over the etched area. The length of vectors corresponds to its absolute value. The bubble #23 has the velocity of 5.3 mm/s.

One can see that near the surface of the liquid the lateral size of bubbles-rich area is three time narrower than the size of the laser-treated zone on the target from which the bubbles are emitted.

The top view of the etched Al target shows the formation of patterns made of gas bubbles. These patterns are clearly seen owing to the dark coloration of the laser-treated area of the substrate. For the square-shaped etched area the picture is as follows (Fig. 4):

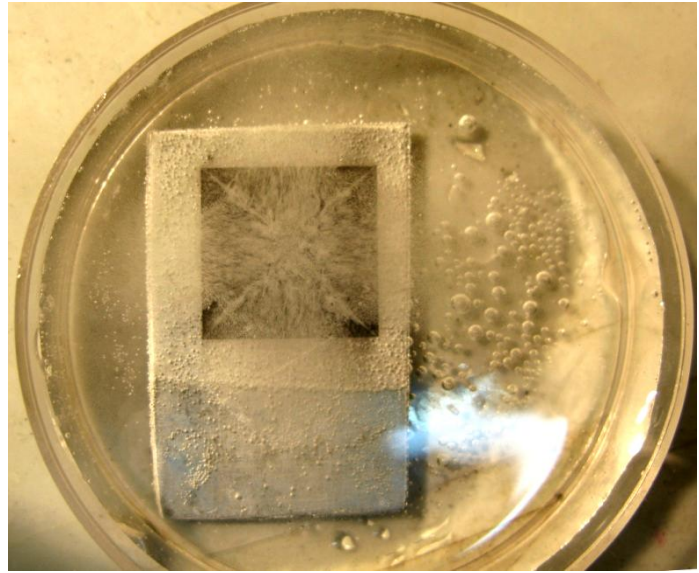

Fig. 4. Stationary pattern of gas bubbles over the square-shaped laser treated area of the Aluminum target. The size of the laser-scanned square is 17x17 mm$^2$. Top view, concentration of $NH_4OH$ is 5%, the height of the solution above the target is around 5 mm.

The gas bubbles are aligned along diagonals of the square. Note that the symmetry of the pattern is completely different of the symmetry of the vessel (circular Petri dish). The pattern formed by gas bubbles is stationary and very stable. It remains the same at least for 2 hours of etching.

The stationary pattern formed by gas bubbles in case of a star-like laser-processed area is presented in Fig. 5. At the beginning of etching the gas bubbles that ascend to the surface of the liquid form a star-like pattern. Then the bubbles are aligned along the rays of the star-shaped etched area. There are very few bubbles on the non-treated area of the target, which facilitates the observation of the pattern above the etched area.

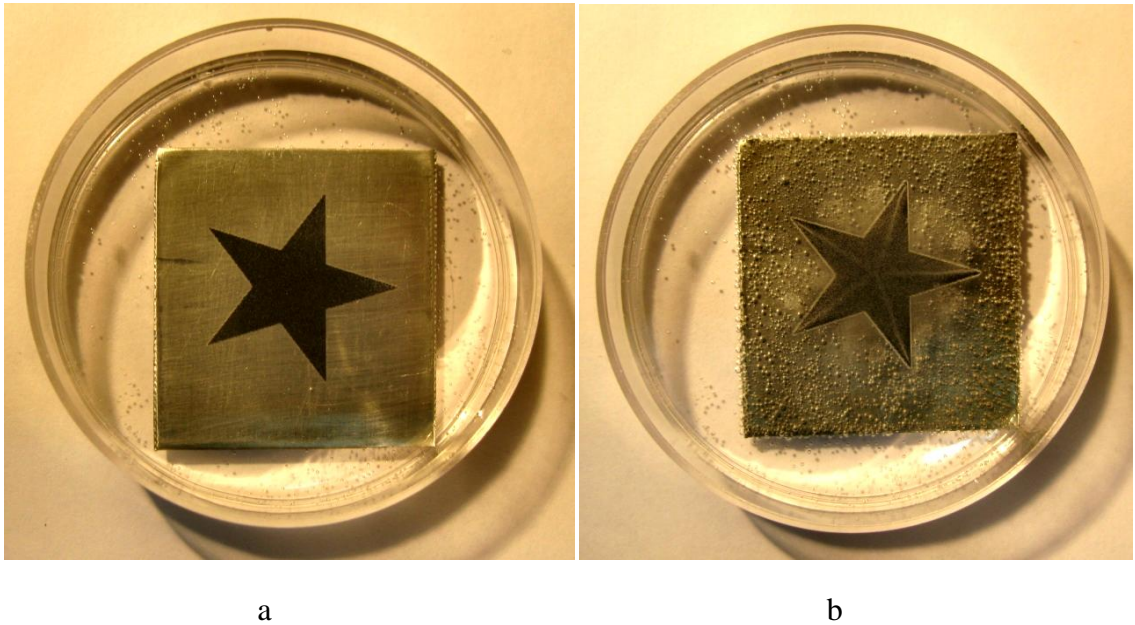

a                                                                                      b

Fig. 5. Formation of the pattern of gas bubbles over a star-shaped laser-processed area. Onset of etching (a), stationary pattern (b). The size of the star-shaped etched area is 15 mm.

Finally, if the etched area has a shape of an arbitrary triangle, the gas bubbles are aligned along the bisectors of the triangles, as demonstrated in Fig. 6.

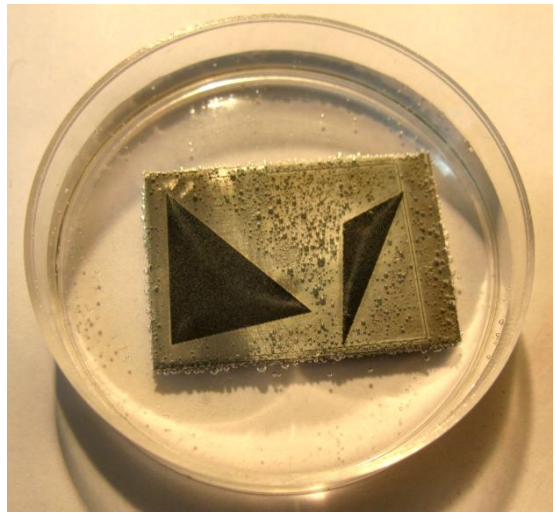

Fig. 6. Stationary pattern of gas bubbles over triangle-shaped areas. The bubbles are aligned along the bisectors of angles.

If the laser-etched area has specific shape shown in Fig. 7, then the entire liquid above the target starts rotating. In this case, however, the thickness of the liquid layer above the target should be at least of several cm. Each bubble has the component of velocity towards the center and also a tangential component according to the curvature of the "petals" of the laser-etched area. For a 4-cm vessel, the linear velocity of the liquid rotation is about several mm/s.

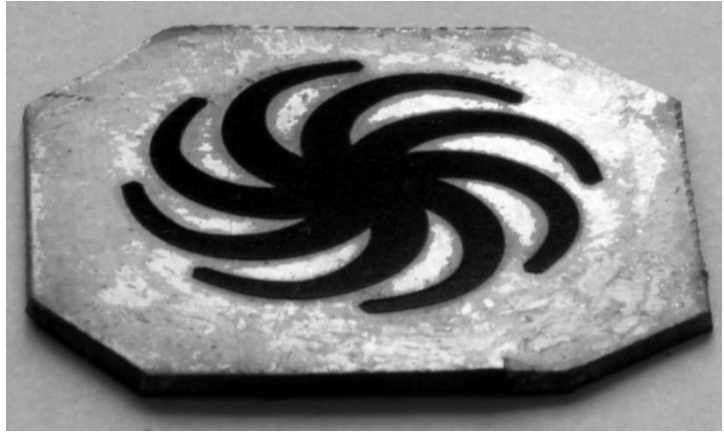

Fig. 7. Macro view of a special vortex pattern of laser-processed area of an Aluminum plate that causes stationary rotation of the liquid in the whole vessel. The diameter of the laser-treated area is 20 mm.

It is clear that the laser here generates only the shape of the preferably etched area. Similar results could be obtained if the shape of the Al plate will coincide with the configuration of the laser-treated area, either in the shape of a square or a star.

**Mathematical simulation**

An individual gas bubble ascends with constant velocity due to the action of two forces, namely, the Stokes viscous force and the buoyancy Archimedes force due to low density of the gas inside it. Each gas bubble engages the surrounding liquid into ascending motion. If the density of bubbles from both sides of the bubble is different, then the velocity of the liquid from both sides of the bubble is also different (Fig. 8). The higher is the bubble density, the higher is the velocity. According to Bernoulli equation, higher velocity corresponds to lower pressure. Therefore, a drift force appears that moves the bubble towards the region of higher velocity of the liquid flow. Therefore, the ascending bubble also drifts to the center of the flow increasing thus the velocity of this flow. This is the positive feedback that governs the formation of a pattern of gas bubbles.

Let us assume that the bubbles are ascending with constant velocity. There is a difference of pressure between the left and right side of a bubble, since the velocity of liquid flow is related to the local density of rising bubbles. This difference of pressure causes the drift of a bubble to the area of higher density of bubbles.

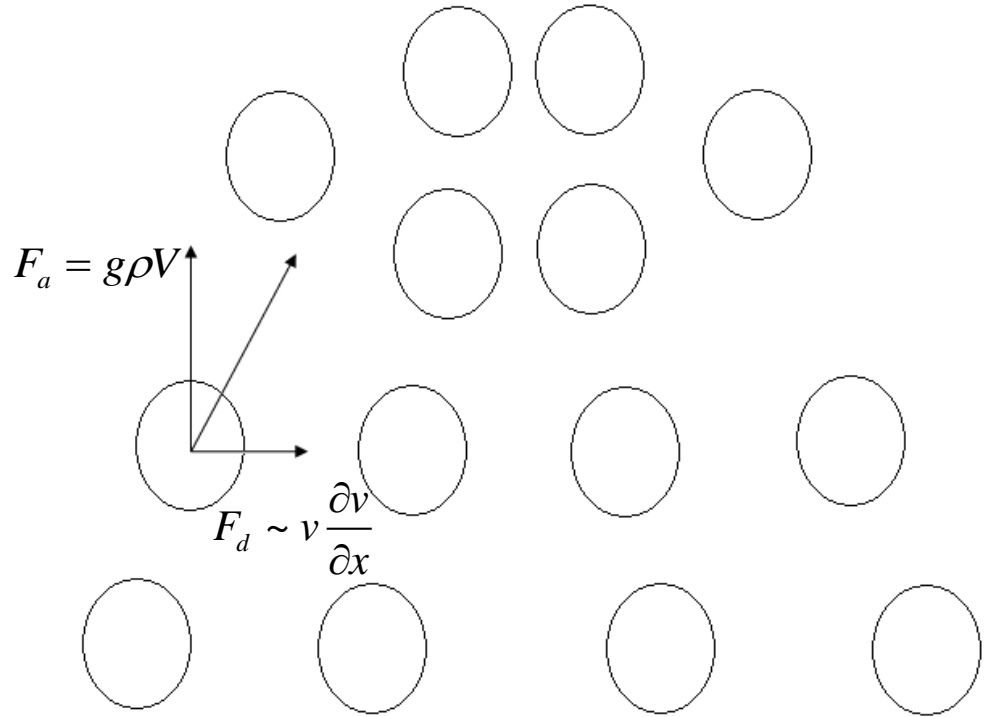

Fig. 8. Forces that determine the bubbles motion: $F_a$ - buoyancy force, $F_d$ - drift force.

The Bernoulli equation for liquid at coordinate x and x+dx is:

$$\frac{mv_x^2(x)}{2} + P(x) = \frac{mv_x^2(x+dx)}{2} + P(x+dx)$$

Here we neglect the variation of pressure with the height of the liquid, which is small in our conditions. Then it follows:

$$-\frac{m}{2} \cdot \frac{\partial v_x^2}{\partial x} = \frac{\partial P}{\partial x}$$

It means that drop of pressure will lead to changes in horizontal component of the velocity. The resulting force is compensated by viscous Stokes.

Let's write on the equation for the distribution function of the bubbles $n(x,t)$ along the horizontal axis. First we should take into account the drift motion of bubbles to the center of the target. For the sake of simplicity we consider that bubbles do not agglomerate with each other and do not expand upon ascending.

The kinetic equation for the function $n(x,t)$ can be written as follows:

$$\partial_t n(x,t) = B \cdot Exp\left(-\frac{(x-x_0)^2}{a^2}\right) - \frac{1}{4} \cdot \partial_x (n(x,t) \cdot v_x(x,t)) - \gamma \cdot n(x,t)$$

$f(x) = B \cdot Exp\left(-\frac{(x-x_0)^2}{a^2}\right)$ describes the formation of bubbles (see Fig. 9). Such curve shape is chosen due to the fact that more bubbles arise in the centre of target than in its edges. $B$ is a constant determined by the characteristics of the chemical reaction (rate of bubbles production).

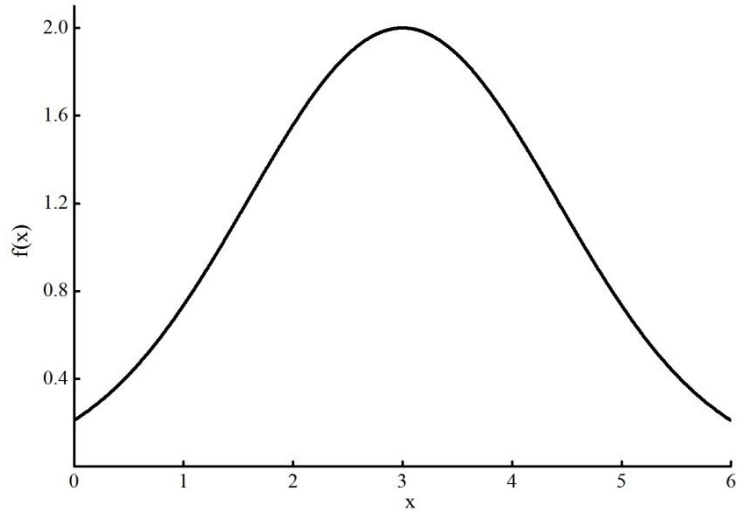

Fig. 9. The Gaussian profile of concentration of bubbles n(x) on coordinate generated in chemical reaction.

The term $\frac{1}{4} \cdot \partial_x (n(x,t) \cdot v_x(x,t))$ describes the drift motion of bubbles in the horizontal plane.

$v_x(x,t)$ is the horizontal component of the bubbles velocity.

The equation for $v_x(x,t)$:

$$\frac{\partial v_x(x,t)}{\partial t} + \alpha \frac{\partial v_x^2(x,t)}{\partial x} = 0,$$

where $\alpha$ is a constant, which contains radius of the bubble. This is Hopfs' equation [1], which can be solved numerically. However we will use simple approximation.

As it follows from the Bernoulli equation, $v_x(x,t) \sim -\sqrt{x-x_0}$ to the right of centre of the target $x_0$. The horizontal component of velocity decreases to $x_0$. In general case

$$v_x(x,t) \sim v_0 \cdot Sign(x_0 - x) \cdot \sqrt{|x - x_0|}$$

For the computation model a function $v_x(x,t) = v_0 \cdot Tg\left(-\dfrac{x - x_0}{a}\right)$ was chosen, where $a$ is the lateral size of the vessel (Fig. 10). We consider a stationary case when the velocity distribution is already established and the number of gas bubbles in the volume remains constant.

$\gamma \cdot n(x,t)$ is a term describing the bubbles escape at the surface of the liquid.

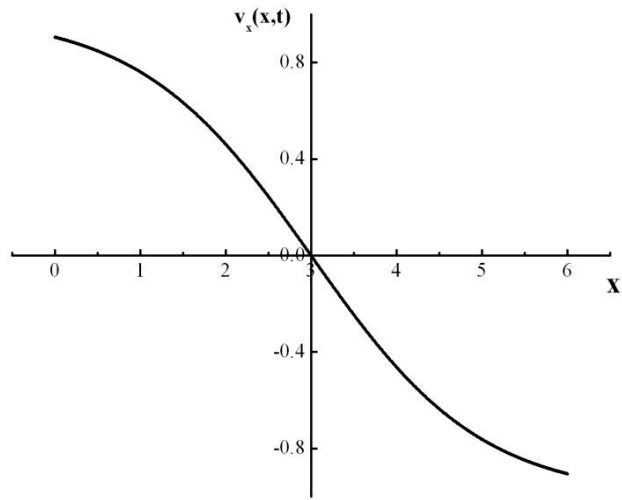

Fig. 10. The dependence of $v_x(x,t)$ on coordinate x.

Boundary and initial conditions are as follows:

$n(x,0) = 0$ - describes the absence of bubbles in initial moment of time: $n(0,t) = 0$ and $n(a,t) = 0$ the number of bubbles in the left and right edge of considered area, $a$ stands for the width of this area.

The results of mathematical simulation are presented in Fig. 11.

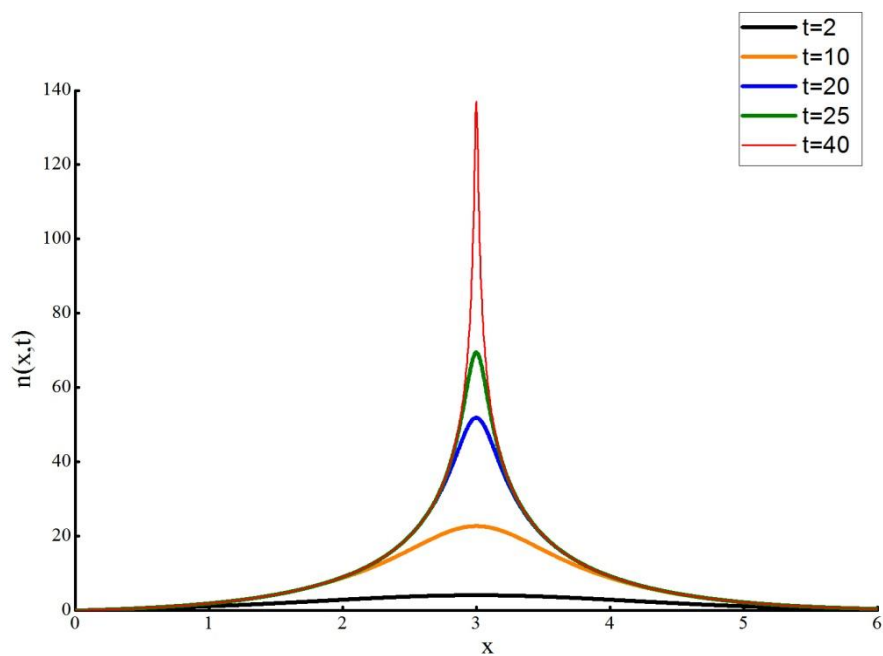

Fig. 11. The dependence of the density of bubbles $n(x,t)$ on coordinate x at different moments of time.

As it follows from the plots, the shape of distribution of bubbles is sharpening with time. This is in good qualitative agreement with the distribution of bubbles shown in Fig. 1. The sharpened profile of bubbles concentration is observed in the experiments over the etched areas of the substrate.

The formation of dissipative structures described in the present work is due to the positive feedback between the bubble motion and liquid flows around it. The source of energy needed for bubbles re-arrangement is the chemical reaction between metal and surrounding solution. Same energy induces rotational movement of the liquid under special profile of the etched surface (Fig. 7).

Thus, a new type of self-organized structures has been described. The structures are formed by gas bubbles that are emitted during chemical reaction of a metal with an ammonium solution. Similar phenomenon is also observed under interaction of Al with either NaOH or KOH, as well as during etching of oxidized Si in HF acid. Such structures should also be observed if the gas is purged into the liquid through an array of micro-apertures.